# Near-field imaging of spin-locked edge states in all-dielectric topological metasurfaces

Alexey Slobozhanyuk[1,2,3], Alena V. Shchelokova[2], Xiang Ni[1,4], S. Hossein Mousavi[5], Daria A. Smirnova[1], Pavel A. Belov[2], Andrea Alù[5], Yuri S. Kivshar[2,3], and Alexander B. Khanikaev[1,2,4]

[1]Department of Electrical Engineering, Grove School of Engineering, The City College of the City University of New York, 140th Street and Convent Avenue, New York, NY 10031, USA

[2]Department of Nanophotonics and Metamaterials, ITMO University, St. Petersburg 197101, Russia

[3]Nonlinear Physics Centre, Australian National University, Canberra ACT 2601, Australia

[4]Graduate Center of the City University of New York, New York, NY 10016, USA

[5]Microelectronics Research Centre, Cockrell School of Engineering, University of Texas at Austin, Austin, TX 78758 USA

E-mail: khanikaev@gmail.com

**A new class of phenomena stemming from topological states of quantum matter has recently found a variety of analogies in classical systems [1,2,3,4,5,6,7,8,9,10]. Spin-locking and one-way propagation have been shown to drastically alter our view on scattering of electromagnetic waves, thus offering an unprecedented robustness to defects and disorder [11,12,13,14,15,16,17,18,19,20]. Despite these successes, bringing these new ideas to practical grounds meets a number of serious limitations. In photonics, when it is crucial to implement topological photonic devices on a chip [21], two major challenges are associated with electromagnetic dissipation into heat and out-of-plane radiation into free space. Both these mechanisms may destroy the topological state and seriously affect the device performance. Here we experimentally demonstrate that the topological order for light can be implemented in all-dielectric on-chip prototype metasurfaces, which mitigate the effect of Ohmic losses by using exclusively dielectric materials, and reveal that coupling of the system to the radiative continuum does not affect the topological properties. Spin-Hall effect of light for spin-polarized topological edge states is revealed through near-field spectroscopy measurements.**

The experimental realization of photonic topological systems have demonstrated fascinating properties for their electromagnetic modes and opened avenues for their use in practical systems and devices [11,15,14,22,23,24,25,26,27]. Unfortunately, many of these original proposals have been based on metal containing metamaterials, [22,28,29,30,26] whose functionality cannot be extended into IR and visible domains. In this respect, all-dielectric topological structures based on arrays of ring resonators [14,31,32] or chiral waveguides [15,27,33] have demonstrated desirable properties at optical frequencies. Nonetheless, this comes at the cost of a very large footprint, with arrays of resonators that can be hundreds to thousands of wavelengths in size, rendering such systems rather unpractical. It is of practical relevance to realize topological systems not only made of photonic compatible materials, but also compact enough for integration with modern photonics circuitry. In this case one has to consider structures with characteristic length scales that are smaller [34] or comparable in size [35] with the wavelength of light. In this regard all-dielectric metamaterials and metasurfaces [36,37,38] recently emerged in the field represent a well-suited platform to realize compact topological photonic insulators. Based on this concept, here we implement a two-dimensional all-dielectric topological metasurface and demonstrate topologically robust photonic transport, and directly visualize the robustness of topological edge states and their spin-locking properties by near field spectroscopy.

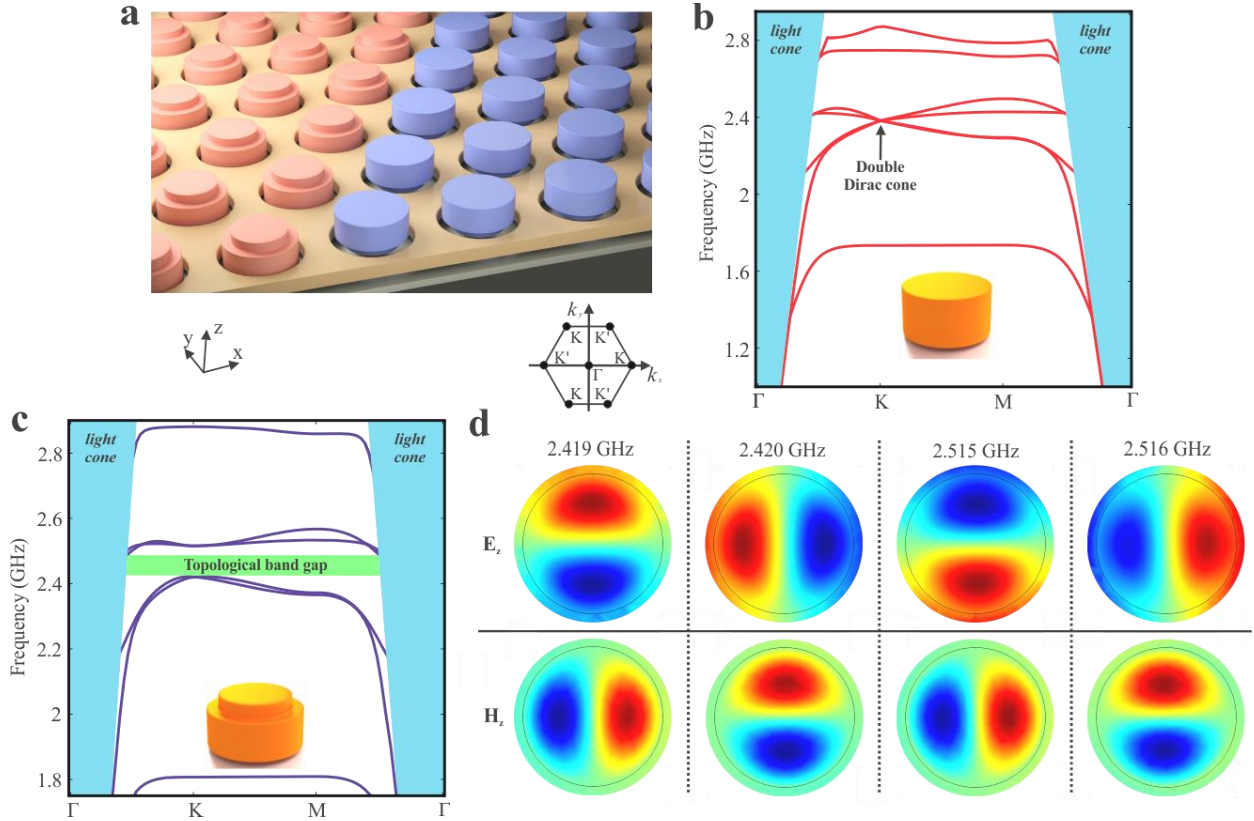

**Figure 1 | Transition from topologically trivial to non-trivial all-dielectric metasurfaces**. (**a**) Geometry of a topologically nontrivial metasurface with a domain wall between arrays that possess different topological characteristics. The period is 36 mm, large disk radius and height are 14.57 mm and 9 mm, small disk radius and height are 11 mm and 3 mm respectively. (**b**) Photonic band structure of the metasurface without bianisotropy. The right panel shows the eigenmodes of the structure near the K-point. (**c**) Photonic band structure of the metasurface with bianisotropy. Insets in panels (b,c) show the corresponding meta-atoms of both metasurfaces. The green shaded area illustrates the spectral width of the topological band gap. The position of the light line is marked by a blue shaded area. (**d**) Field profiles of eigenmodes of the bianisotropic metasurface, revealing locking of their electric and magnetic field components.

The structure that we study is schematically shown in Fig. 1(a) and it represents a triangular array of dielectric disks with their dimensions tuned such that the array supports two doubly degenerate Dirac cones at K and K' points of the Brillouin zone. The band structure plotted in Fig. 1(b) reveals two pairs of Dirac bands, stemming from electric and magnetic dipolar modes of the disks, respectively, and their degeneracy effectively restores the dual symmetry responsible for the topological properties we exploit here [17,28,26,39]. The field configurations for the corresponding dipolar modes carrying angular momentum $l = \pm 1$ are shown in Fig. 1b (as polarized in x- and y-direction) and the presence of Dirac points stems from the degeneracy of these modes at K and K' points due to PT and the rotational symmetries of the lattice.

The double degeneracy between electric and magnetic dipolar modes enables emulating the spin degree of freedom of topological insulators [39,40]. On the other hand, coupling between electric and magnetic modes, referred to as magneto-electric coupling (or bianisotropic response), allows emulating coupling of spin and orbital degrees of freedom, essentially the spin-orbit interaction of light [17,39]. To induce such coupling, we use the fact that electric and magnetic modes have different parity in the out-of-plane (z) direction, and break this symmetry by introducing a circular notch on one of the flat faces of the cylinders. This gives rise to hybridization of electric and magnetic dipolar modes and opening a complete photonic band gap (Fig. 1c). Note that, while this gap is complete below the light line, eigenmodes located above the light line may couple to the continuum of radiation in the cladding due to the finite thickness of the structure. Nonetheless, the top and bottom bands do not cross anywhere in the Brillouin zone, which implies that the topological properties are completely defined by the hybridization of magnetic and electric Dirac bands near the K and K' points.

The eigenmodes of the structure with bianisotropy are mixed states with in-plane electric and magnetic dipolar components being locked in both orientation and phase (Fig. 1d). These mixed states are referred to as pseudo spin-up (↑) and spin-down (↓) states and appear pairwise as top and bottom bands in the spectrum. To confirm that these new states possess topological properties, we performed numerical calculations of the topological invariant [28,29], and found that the upper and lower modes are characterized by non-vanishing spin-Chern numbers $C_{l/u}^{\uparrow} = \pm 1$ and $C_{l/u}^{\downarrow} = \mp 1$, where the subscripts indicate lower (l) and upper (u) bands. Importantly, the reversal of the orientation of the meta-atoms in the vertical direction (from the notch up to the notch down) leads to the reversal of the pseudo-spins and of the bands in the spectrum, which is also accompanied by an inversion of the topological invariant to $C_{l/u}^{\uparrow} = \mp 1$ and $C_{l/u}^{\downarrow} = \pm 1$, respectively. This fact is subsequently used to create topological interfaces.

The proposed structure was fabricated using high index ($\epsilon = 39$) ceramic discs of two radii, which were glued one to another to from bianisotropic meta-atoms. Each of such meta-atom has been characterized independently by measuring their individual response to confirm desirable characteristic in the frequency range of interest (see Methods). As the next step, the particles have been arranged into a triangular array and the collective response was studied by RF spectroscopy. First, we confirmed that the transmission spectra of array possess a bandgap induced by the bianisotropy in the frequency range 2.42-2.52 GHz (Fig. 2a). Our measurement results are in full agreement with the numerical calculations of the band structure. Figure 2b(1) shows that the modes excited by the dipolar source placed inside the crystal are completely localized near the source, e.g. are not propagating towards any direction.

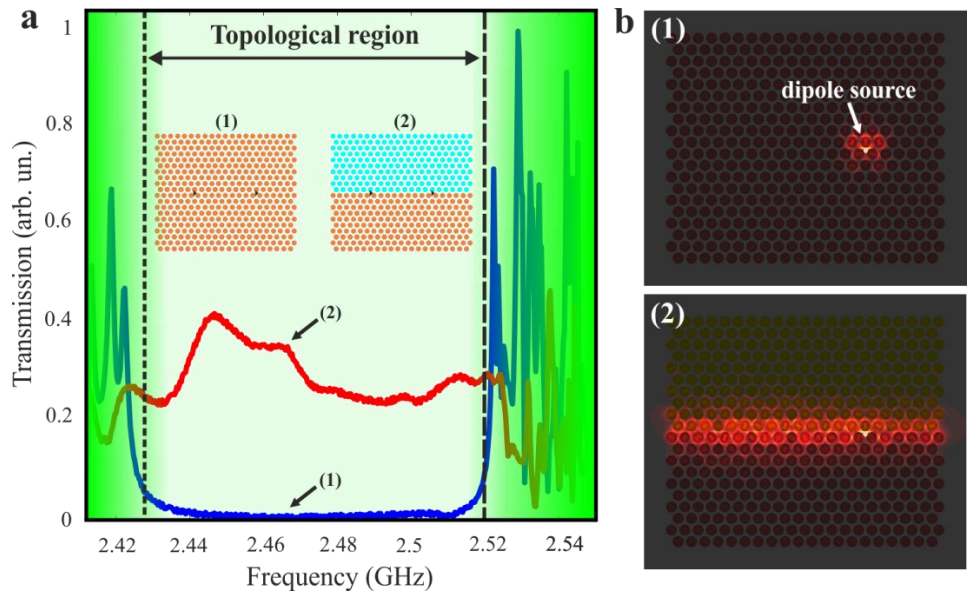

**Figure 2 | Spectral characteristics of all-dielectric topological metasurfaces.** (**a**) Experimental transmission spectra of the topological metasurface without domain wall (blue line) and along the domain wall (red line). Location of the dipolar sources is indicated by the black triangles in the left inset. The insets schematically show the geometrical configurations used in both experiments. (**b**) Simulated distribution of the electric field amplitude at the frequency of 2.468 GHz for two different configurations of the metasurfaces, without the domain wall (1) and with the domain wall (2) which supports the topological edge states.

The major interest in topological electromagnetic phases lies in the fact that they support robust edge states at their interfaces, thus enabling guiding around defects. The interfaces supporting such modes can be divided into two classes: i) interfaces with non-topological domains and ii) domain walls, which represent interfaces between structures possessing different topological characteristics. Since the topological order originates from the dual symmetry of the structure, we have to limit our study to interfaces that do not violate such the symmetry. On the other hand, the external boundary with free space may appear a good symmetry-protecting interface. However, it does not allow confining the topological states, due to strong radiative coupling into the electromagnetic continuum. We therefore choose the second class of a topological interface, a domain wall separating two topologically distinct domains of the metasurface. The two domains represent crystals with opposite orientation of the meta-atoms in the vertical direction, thus effectively creating a topological interface, the domain wall, across which the spin-Chern numbers experience a change of magnitude 2. The bulk-interface correspondence [17] implies that such a domain wall should host a total of four states, two with spin-up and two with spin-down, respectively.

The numerical simulations carried out with the use of COMSOL Multiphysics for a supercell of 24 meta-atoms with a domain wall (the flip of orientation from up to down) in the middle and the perfectly matching boundary condition applied in vertical direction at the top and bottom of the structure confirm the existence of edge states. The edge states appear within the topological bandgap and are strongly localized to the domain wall in the lateral direction, and while they do lay below the light cone, they apparently overlap spectrally with the modes of radiative continuum laying above the light line (Fig. 3a). Despite this, the edge modes appear to be well defined and strongly localized to the domain wall and, as demonstrated below, have negligible leakage to the continuum. Importantly, no such localized edge modes have been found for similar calculations carried out for the case of all meta-atoms facing up, while an identical bulk spectrum was obtained (not shown), thus confirming the topological origin of the edge states.

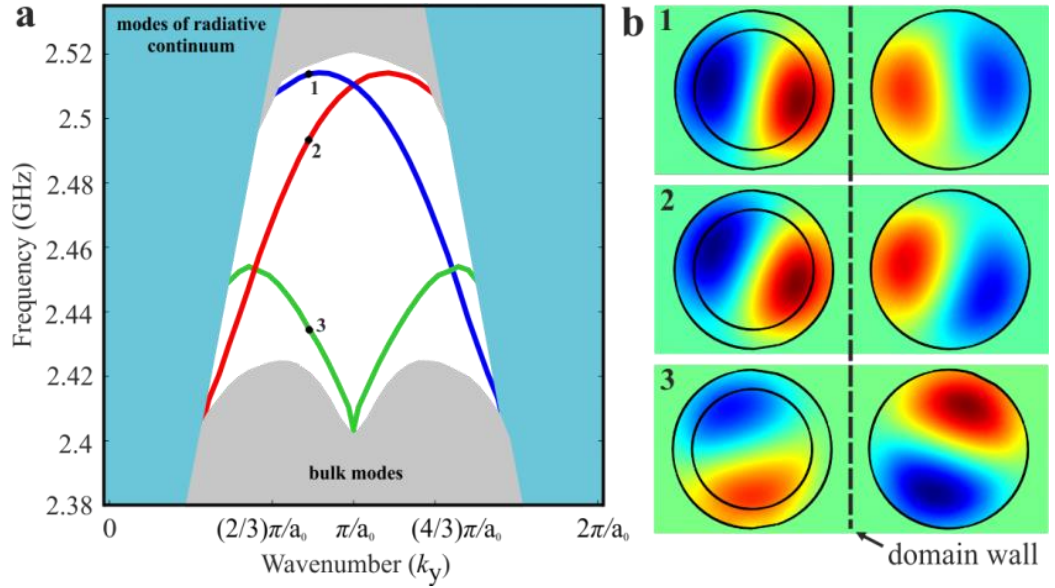

**Figure 3 | Topological edge modes supported by the domain wall in open all-dielectric metasurface.** (a) Dispersion of the photonic topological edge states supported by the domain wall in the metasurface. The gray shaded areas illustrate the bulk photonic states, while the blue shaded area corresponds to the modes of radiative continuum. (b) The field profiles ($D_z$) across the domain wall for different frequencies, corresponding to the different edge modes, indicated in the panel (a).

Interestingly, only two of the modes appear to cross the entire topological band gap, while the second pair touches the light line and disappears in the middle of the gap after emerging from the continuum of the bottom bulk modes. This is important distinction of the open metasurface from its closed photonic counterparts in which 4 topological edge modes would also cross the entire bandgap due to the absence of the light cone and the continuum of modes within it [17, 29, 26]. The inspection of the field profiles for the two sets of modes also shows that the ones crossing the gap represent a set of even modes symmetric across the domain wall (Fig. 3b). For these modes, one spin-up and one spin-down, respectively, the coupling to radiative continuum is largely suppressed by their symmetry, which ensures that such coupling does not destroy the topological nature. Such suppressed coupling can be easily understood by noticing that these modes have quadrupolar $Q_{x^2-y^2}$ profile which has very low radiation efficiency. Meanwhile, the odd edge states not only cross the light cone, but appear to interact strongly with the radiative continuum which affects their dispersion by pushing bands to the low frequency range - a clear indication of destruction of their topological properties caused by the interaction with the continuum. This strong interaction of the odd edge modes with the continuum can also be intuitively understood from their field profile across the domain wall, which has clear out-of-plane magnetic dipolar $M_z$ (and quadrupolar $Q_{xy}$) configuration and efficiently radiates in the directions close to lateral.

To experimentally confirm the existence, spin-locked propagation, and robustness of the topological edge states, we changed the configuration of the array such that half of the meta-atoms were oriented upward and the second half – downward thus forming the domain wall across the straight line (Fig. 4a). The edge modes were then excited by a dipole antenna placed in the middle of the domain wall in the horizontal direction and just above the array to ensure coupling to the near field (see the right inset in Figure 2 and Methods for details). Next, the presence of the topological edge mode was tested by placing the dipole source at the domain wall and the transmission spectrum was measured (see Figure 2). As expected, enhanced transmission within the band gap region occurred due to the excitation of the edge mode, as shown in Fig. 2a red curve and the lower panel of Fig. 2b. The precise location of the antenna allowed us exciting the topological edge states with electric field maximum at the center of the domain wall. At the same time, controlling the polarization of the dipole antenna enables selective excitation of the mode with specific angular momentum. Thus, circular left or right polarization excites one of the two particular pseudo-spins due to the locking of spin and angular momentum degrees of freedom. As it can be seen from the map of the near field collected by another antenna attached to the xy-motorized stage, the linear polarization excites edge modes of both spins running in two opposite directions away from the source, respectively (Fig. 4b). However, switching to circular polarizations changes the pattern of the field and, as shown in Fig. (4c), the left and right handed excitation, and results in a unidirectional flow of the radiation from the source. These experimental results unambiguously confirm the spin-locking and the topological character of the edge states.

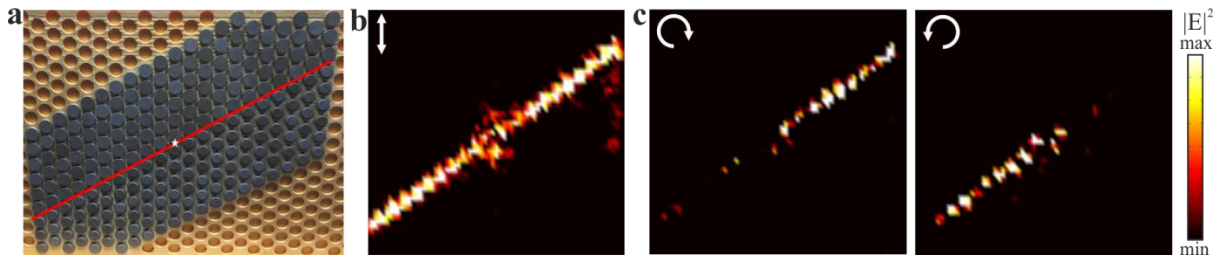

**Figure 4 | Fabricated prototype of all-dielectric topological metasurface and experimental observation of the spin-locking and one-way excitation**. (**a**) Photograph of the metasurface with the location of the domain wall indicated by the red line and the source position marked by the star symbol. (**b,c**) Experimentally measured electric near-field intensity images for the linear (b) right- and left- circular (c) polarizations of the source at the frequency of 2.462 GHz.

The observed spin-locking is known to be responsible for the unprecedented robustness of the edge states to structural imperfections and defects, which has been demonstrated before for metamaterial systems based on metal components [22, 28, 29, 26]. Here we expect to see similar robustness in an all-dielectric platform, while the open geometry of the system offers additional benefits by allowing us to directly visualize reflectionless flow of the field by the near-field measurements. To confirm such robustness, we created a domain wall configuration with two sharp bends, and carried out experimental measurements with a source placed in the middle of the central segment, as shown in Fig. 5a. The circularly polarized feed again leads to selective excitation of spin-up or spin-down modes, which flow towards a particular bend. As expected from the topological robustness, the modes flow around the bends without back-reflection, as confirmed by the near-field map in Fig. 5a (right panel). Similar unidirectional excitation and robust propagation was also found for the case of the source placed in the first long arm of the domain wall (Fig. 5b), in which case the spin-up edge state propagates across two sharp bends without back reflection and avoided the formation of standing-wave field-configurations in the central segment.

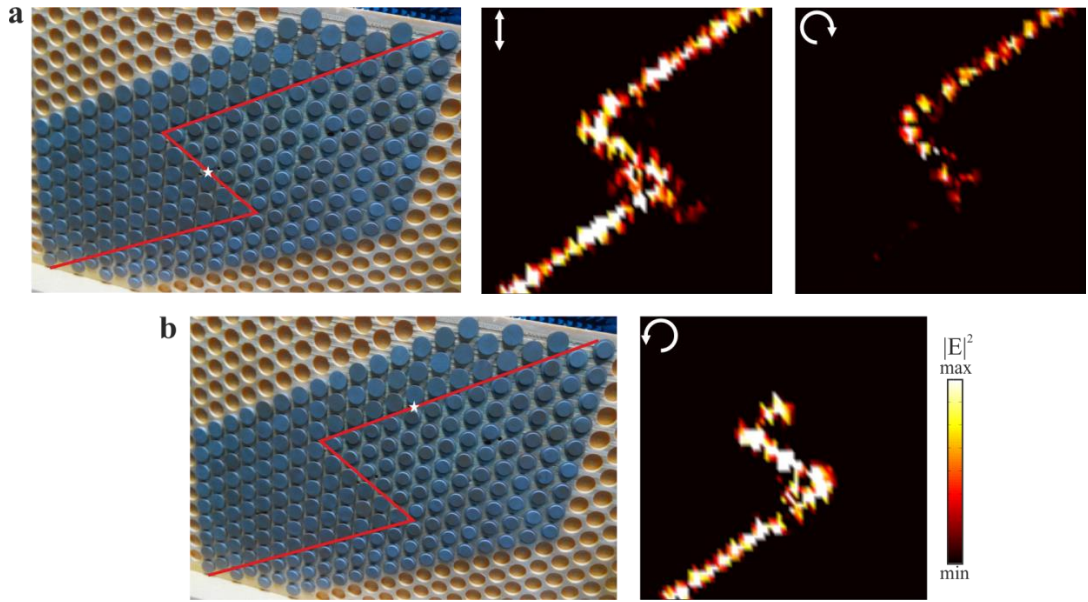

**Figure 5** | **Experimental verification of the robustness of the edge state against sharp bends**. (**a**) (Left panel) Photograph of the metasurface with the location of the double-bend domain wall indicated by the red line and the source position marked by the star symbol. (Right panels) Experimentally measured electric near-field intensity images for the linear and right-circular polarizations of the source at the frequency of 2.462 GHz. (**b**) (Left panel) Photograph of the metasurface that indicates the displacement of the source (marked by the star symbol) together with (right panel) the electric near-field intensity image for the left-circular polarization of the source at the frequency of 2.462 GHz.

To summarize, a photonic topological metasurface has been experimentally implemented using all-dielectric materials and tested in the microwave frequency range. Despite the open nature of the system, we found that the system maintains its bulk topological characteristics. The topological interfaces in the form of domain walls were tested and found to host two types of edge states, with only one of them interacting and coupling with the continuum of modes above the light line. As a result, this type of edge state loses its topological characteristics merging with the radiative continuum without crossing the topological bandgap. The second class of edge states is unaffected by the presence of the radiative continuum and exhibits topological properties such as spin-locking and lacks back scattering from defects. The open nature of the system allowed direct visualization of these properties, unambiguously confirming the topological robustness of electromagnetic transport along the domain wall.

We believe that the proposed all-dielectric metasurface platform opens a new avenue for advanced photonic technologies across the entire electromagnetic spectrum, and it brings topological photonics closer to practical use. In particular, the use of dielectric materials and the small footprint of the proposed design makes it compatible with existing semiconductor devices, which envisions integration of photonic and electronic components on the same substrate. At the same time, the topological nature of the design provides robust transport and versatile control of optical signals by synthetic gauge fields.

**Methods**:

*Numerical simulations:* The commercially available full-wave electromagnetic simulator software COMSOL Multiphysics (version 5.0) and CST Microwave Studio 2016, were used to generate all numerical simulations presented in the figures of the main text. Specifically, we carried out eigenfrequency analyses in COMSOL in order to calculate photonic bulk and edge band structure of the metasurfaces and field profiles. For unit cell simulations (Figure 1) we have used periodic boundary conditions in *x*- and *y*-directions and PML in *z*-direction. For the supercell simulation (Figure 3), we have used the same boundary as before for 24 meta-atoms in order to be able to design the domain wall. Full wave simulations of the finite metasurfaces prototypes were carried out with the aid of frequency domain solver of CST Microwave Studio. We have used PML boundary conditions and added additional space between the structure and the boundaries in order to avoid possible minor reflections. The excitation was performed via the insertion of the subwavelength dipole inside the structure.

*Experiments:* The metasurface structure was fabricated using high permittivity ($\epsilon=39$) ceramic discs of two radii which were glued one to another. The spectral response of each meta-atom has been measured in the frequency range of interest, and the positions of the resonances have been determined via an inspection of the reflection coefficient of the small loop antenna (diameter 11 mm) placed above the meta-atom. The metasurface was formed through the insertion of the meta-atoms in the machine-processed substrate made from Styrofoam material (epsilon around 1.2 and negligible losses), where the holes with certain diameters and periodicity were drilled. All the measurements were performed in the anechoic chamber. We employed two subwavelength dipoles as a source and a receiver, which were connected to the vector network analyzer (Agilent E8362C) for measurement of the transmission spectra shown in Figure 2. To perform the near-field measurements, we used an automatic mechanical near-field scanning setup and an electric field probe connected to the receiving port of the analyzer (Figures 4, 5). The probe was normally oriented with respect to the interface of the structure. The near field was scanned at the 2 mm distance from the back interface of the metasurface to avoid a direct contact between the probe and the sample. In order to determine the structure response to the right- and left-circular polarizations of the source, we collect the amplitude and phase of the field excited near the structures by a linearly polarized dipole oriented along *x*- and *y*-directions. The resulting experimental maps were obtained by combining the measured complex fields for two orthogonal excitation polarizations in MATLAB.

**Acknowledgements**

The authors are grateful to E. Nenasheva for the fabrication of ceramic resonators and D. Filonov for an assistance with experimental measurements at the initial stage of this project. The work was supported by the National Science Foundation (grants CMMI-1537294 and EFRI-1641069). The experimental part of the work was supported by the Russian Science Foundation (grant no.16-19-10538). The work of YSK and AS was partially supported by the Australian Research Council. Research was partly carried out at the Center for Functional Nanomaterials, Brookhaven National Laboratory, which is supported by the US Department of Energy, Office of Basic Energy Sciences, under contract no. DE-SC0012704.

**Author contributions**

All authors contributed extensively to the work presented in this paper.

**Competing financial interests**

The authors declare no competing financial interests.